# Iron snow in the Martian Core?

Christopher J. Davies[1,2,*] and Anne Pommier[2]

[1]School of Earth & Environment, University of Leeds, Leeds LS2 9JT, UK.

[2]Institute of Geophysics and Planetary Physics, Scripps Institution of Oceanography,

University of California at San Diego, 9500 Gilman Drive, La Jolla, CA 92093-0225, USA.

*Correspondence to: c.davies@leeds.ac.uk.

**Abstract**

The decline of Mars' global magnetic field some 3.8-4.1 billion years ago is thought to reflect the demise of the dynamo that operated in its liquid core. The dynamo was probably powered by planetary cooling and so its termination is intimately tied to the thermochemical evolution and present-day physical state of the Martian core. Bottom-up growth of a solid inner core, the crystallization regime for Earth's core, has been found to produce a long-lived dynamo leading to the suggestion that the Martian core remains entirely liquid to this day. Motivated by the experimentally-determined increase in the Fe-S liquidus temperature with decreasing pressure at Martian core conditions, we investigate whether Mars' core could crystallize from the top down. We focus on the "iron snow" regime, where newly-formed solid consists of pure Fe and is therefore heavier than the liquid. We derive global energy and entropy equations that describe the long-timescale thermal and magnetic history of the core from a general theory for two-phase, two-component liquid mixtures, assuming that the snow zone is in phase equilibrium and that all solid falls out of the layer and remelts at each timestep. Formation of snow zones occurs for a wide range of interior and thermal properties and depends critically on the initial sulfur concentration, $\xi_0$. Release of gravitational energy and latent heat during growth of the snow zone do not generate sufficient entropy to restart the dynamo unless the snow zone occupies at least 400 km of the core. Snow zones can be $1.5 - 2$ Gyrs old, though thermal stratification of the uppermost core, not included in our model, likely delays onset. Models that match the available magnetic and geodetic constraints have $\xi_0 \approx 10\%$ and snow zones that occupy approximately the top $100$ km of the present-day Martian core.

## 1. Introduction

Low-altitude vector magnetometer measurements from Mars Global Surveyor show that Mars presently lacks a global dipole field, but reveal large regions of strongly magnetized crust located mainly in the southern highlands (Acuña et al. 1998). The prevailing view is that this magnetization was acquired as the rock cooled in the presence of a global magnetic field (Stevenson, 2001; Breuer and Moore, 2015). The global field was likely produced in the liquid core by a dynamo process in which thermal (and possibly chemical) buoyancy forces drive convective motion (Stevenson, 2001). Inferences based on the age of impact craters (Acuña et al., 1998; Langlais et al., 2012) and Martian meteorites (Weiss et al., 2002) suggest that the global field decayed around 3.8-4.1 Ga. This event marks the demise of the Martian dynamo and may have been contemporaneous with changes in the planets' heat loss (Ruiz, 2014) and oxidation state (Tuff et al., 2013).

Explanations of Mars' magnetic history are intimately linked to the thermal evolution and crystallization regime of its metallic core. A thermal dynamo can operate in an entirely liquid core, provided that the ancient core-mantle boundary (CMB) heat flow $Q_{cmb}$ exceeded the heat $Q_a$ lost by conduction down the adiabatic temperature gradient (assuming no radiogenic heating). In this scenario the core cooled, perhaps from an initially superheated state compared to the mantle (Williams and Nimmo 2004) or modulated by an early episode of plate tectonics (Nimmo and Stevenson, 2000), until $Q_{cmb}$ fell below $Q_a$. Impact-induced thermal insulation of the core (Monteux et al., 2013; Arkani-Hamed and Olson, 2010) would produce a similar outcome. On the other hand, a thermochemical dynamo can operate with $Q_{cmb} < Q_a$. It has been suggested that rapid growth of an inner core early in Mars' history led to dynamo termination when the size of the liquid region fell below a critical threshold (Stevenson, 2001). This scenario has not been favored because inner core growth provides additional power sources that lead to a long-lived dynamo (Williams and Nimmo, 2004). In this paper we investigate a third scenario: the top-down crystallization of the Martian core.

A necessary condition for top-down core freezing is $\partial T_l/\partial P < \partial T/\partial P$ for all pressures $P$, where $T_l$ is the liquidus temperature of the core alloy and $T$ is the ambient temperature. $\partial T/\partial P$ is positive and for an adiabatic core $\partial T_a/\partial P \propto T_a \propto T_{cmb}$, where $T_{cmb}$ is the CMB temperature. Melting curves for iron-sulfur systems, the mixture used throughout this work, have been extensively studied (Kamada et al., (2012); Morard et al., (2011); Figure 2d). Of particular interest are the results of Stewart *et al.*, (2007) who found that $\partial T_l/\partial P < 0$ at the $P-T$ conditions of Mars' core using Fe-S mixtures with 10.6 $wt\%$ and 14.2 $wt\%$ S, which assures top-down cooling over bottom-up cooling. However, application of top-down crystallization to Mars depends critically on whether its core has cooled sufficiently over the last 4.5 billion years for $T_{cmb}$ to fall below $T_l$. A further issue is that additional power sources accompanying top-down crystallization could have provided sufficient entropy to restart the dynamo.

Here we build a parameterized model of top-down crystallization in the Martian core. We consider the so-called “iron snow” regime that arises when the bulk sulfur concentration is smaller than the sulfur concentration of the eutectic composition: solid produced on freezing is heavier than the residual liquid and iron “snows” down onto the underlying liquid (Hauck et al., 2006; Dumberry and Rivoldini, 2015; Rückriemen et al., 2015). We follow the premise of previous work (Hauck et al., 2006; Rückriemen et al., 2015) and assume that crystallization in the snow zone produces a slurry: solid particles are suspended in a liquid Fe-S mixture and the solid fraction $\phi$ remains low enough that the system behaves as a liquid. The fluid dynamical behavior of a binary slurry is fundamentally different from that of a binary liquid mixture and so the theory must be developed from scratch, starting with the fundamental conservation equations. We derive energy and entropy equations from an established slurry theory (Loper and Roberts 1977) that does not appear to have been utilized in previous models of iron snow in planetary cores.

Our model assumes that the snow layer is always in phase equilibrium and that freezing produces iron solid that quickly falls to the deeper liquid core (see Sections 2 and 4 for detailed discussion of the modelling approximations). Starting from an equilibrium state with the entire region at the liquidus temperature, cooling reduces $T$ below $T_l$ leading to the formation of solid iron (Figure 1). The local increase in $\phi$ releases latent heat and elevates the sulfur concentration in the coexisting liquid phase, which in turn depresses $T_l$ until it reaches $T$ ($T_l < T$ implies the layer is fully liquid). Assuming no net mass exchange between core and mantle on the timescales of interest, the light residual liquid rises, producing a stable chemical stratification across the snow zone (Dumberry and Rivoldini, 2015; Rückriemen et al., 2015). The heavy solid sinks out of the snow layer into the underlying liquid region where it remelts, absorbing latent heat and causing a decrease in sulfur concentration and an increase in $T_l$. Gravitational energy is liberated in the snow zone due to iron sinking and also in the liquid due to stirring induced by dense iron remelting (Rückriemen et al., 2015; Figure 1). These additional heat sources, together with variations in composition and temperature across the snow zone induced by freezing, influence the core cooling rate and the power available to generate a magnetic field.

The complexity of the iron snow equations together with uncertainties in thermal and material properties of Fe-S alloys at high $P - T$ conditions mean that we do not expect (or attempt) to obtain a definitive thermal history for Mars. Rather, we seek to understand the conditions that could have led to snow zone formation. Nevertheless, viable models must be compatible with the magnetic history of Mars and with geodetic observations, which suggest that at least part of the Martian core is liquid at the present day (Yoder et al., 2003).

## 2. Model and Methods

We generalize an existing 1D thermochemical evolution model (Davies, 2015) to study crystallization of an iron-sulfur alloy (Taylor, 2013) in the Martian core. A standard averaging procedure (Nimmo, 2015) is used to obtained the equations governing changes in the *reference* or *equilibrium* state, in which variables depend only on radius $r$. In regions where there is no solid and outside thin boundary layers it is assumed that vigorous convection maintains a reference state where the pressure $P$ is determined by a hydrostatic balance, the sulfur concentration $\xi$ is uniform and the radial entropy gradient is zero (Braginsky and Roberts 1995). These conditions imply that temperature follows an adiabatic profile.

The global energy budget determines the core evolution by balancing $Q_{cmb}$ against the sum of the heat sources within the core as defined below. The energy balance does not contain information about the dynamo because magnetic energy is converted to heat within the core. The entropy balance contains the irreversible processes of thermal, chemical, mechanical and Ohmic dissipation. Together, these equations describe the thermal and magnetic history of the core.

The general slurry theory describes the time-dependence of particle composition and local departures from phase equilibrium (Loper and Roberts, 1977) and must be simplified for application to planetary cores. We therefore adopt the following two approximations espoused by Loper and Roberts (1977): 1) No light element partitions into the solid phase on freezing; 2) “fast melting”, i.e. instantaneous relaxation to phase equilibrium. The first approximation is supported by experiments that reported very low sulfur concentrations in the solid phase (Kamada et al., 2012; Li et al., 2001) and has also been used to model iron snow in Ganymede’s core (Rückriemen et al., 2015). The second approximation means that material in an infinitesimal volume of the continuum will melt/freeze instantly.

In an equilibrium snow zone, the entire system is on the liquidus and the liquidus is collinear with the adiabat. Heavy sulfur-depleted solid sinks and eventually falls out of the layer where it remelts because the temperature of the underlying liquid is above the liquidus. We assume, as in Rückriemen et al. (2015), that the timescale for sinking and remelting is much faster than the timescale for changes to the equilibrium state. At each timestep, all of the newly created solid sinks out of the layer and remelts, leaving the layer on the liquidus. We refer to this third approximation as “fast remelting”.

The temperature profile may not be adiabatic throughout the Martian core because compositional and/or thermal stratification can develop below the CMB. Consider first the case where $Q_{cmb} > Q_a$, i.e. the temperature profile is everywhere unstably stratified. Subsequent growth of a snow zone will produce a stable compositional stratification below the CMB. In this case it can be shown using equation (8) below that the isentropic condition requires

$$\frac{dT}{dr} = -\frac{\alpha g T}{C_p} - \frac{T\bar{s}}{C_p}\frac{d\xi}{dr}$$

where $C_p$ is the specific heat capacity, $\bar{s} > 0$ is the heat of reaction coefficient defined below, and the fast remelting approximation and $\phi \ll 1$ have been used to remove the contribution from radial variation in solid fraction.  The first term on the right-hand side is the usual definition of the adiabatic temperature in a homogeneous fluid. The second term increases $|dT/dr|$ since $d\xi/dr > 0$ and shows that there must be a greater variation in temperature in the presence of a stabilizing compositional gradient in order to keep the layer isentropic. Accounting for this second term is complicated because $d\xi/dr$ is determined from the liquidus, which is itself related to $dT/dr$ (see below). Instead of undertaking a complex iterative procedure, which seems unnecessary in light of significant uncertainties in several of the model parameters, we ignore variations in $\xi$ in the adiabat. *A posteriori* estimates (Section 4) reveal that this is a good approximation.

The configuration of unstable thermal stratification and stable compositional stratification is potentially susceptible to oscillatory double-diffusive instabilities since heat and mass have different diffusion coefficients (Turner, 1973). In the standard doubly-diffusive configuration the horizontally averaged temperature is not expected to deviate significantly from the original adiabatic profile in this case (Buffett and Seagle, 2010) and any effect on terms in the global energy and entropy budgets should be minor. These results may not apply to an equilibrium slurry where Fickian diffusion no longer holds (Loper and Roberts, 1977); however, in the absence of theoretical or experimental evidence to the contrary we assume that doubly diffusive effects do not influence the adiabatic profile.

If $Q_{cmb} < Q_a$ a region at the top of the core will become stable to thermal convection. The base of the thermally stable layer is located where the stabilizing thermal buoyancy balances the destabilizing buoyancy forces that drive convection in the deeper core (Lister and Buffett, 1998). Departures from an adiabat are expected to be small because the thermal diffusion time, $\tau_d = \delta^2/\kappa$ where $\delta$ is the thickness of the thermally stable layer and $\kappa \approx 10^{-6}$ is the thermal diffusivity, is around $10^7$ yrs even for layers as thin as 10 km, and should not affect estimates of terms in the global equations significantly. Thermal history calculations for the Earth's core with and without a thermally stratified layer showed only minor differences to the global energy balance (Labrosse et al., 1997), which were caused primarily by the assumption that gravitational energy release occurs only in the unstably stratified region rather than by departures from adiabaticity. A more important effect arises because the inability of mantle convection to evacuate all of the heat brought to the CMB by core convection requires that the top of the core must heat up. Since snow zones form when $T_{cmb} < T_l$ at the CMB, formation will be delayed when thermal stratification is present compared to when it is absent. Unfortunately, a thermally stable layer will, in general, not grow at the same rate as a snow layer; creating a parameterization for the dynamics and couplings between regions of different thermal and compositional stability significantly complicates the model and obscures the effects associated with the slurry that we wish to investigate. Our model considers an entirely adiabatic core and will therefore predict a lower $T_{cmb}$ and earlier snow zone nucleation than would be obtained if thermal stratification were taken into account; we return to consider the impact of thermal stratification when applying the results to Mars.

The main assumptions used to develop a quantitative model for the equilibrium evolution of the snow zone are:

1) All sulfur remains in the liquid phase on freezing.

2) Fast melting, i.e. instantaneous relaxation to phase equilibrium.

3) Fast remelting of sinking solid, i.e. rapid sinking and remelting of solid iron.

4) An adiabatic temperature profile exists throughout the core.

Using approximations (1) and (2) the general thermal energy equation in a slurry can be written (Loper and Roberts, 1977)

$$\rho T \frac{Ds}{Dt} = -\nabla \cdot \boldsymbol{q} + \mu \nabla \cdot \boldsymbol{i} + \frac{\mathbf{J}^2}{\sigma}, \tag{1}$$

where the density $\rho$, temperature $T$, entropy $s$, and chemical potential of the liquid $\mu$ are all functions of radius $r$ and $D/Dt$ denotes the material derivative. The heat flux vector $\boldsymbol{q}$ and mass flux vector $\boldsymbol{i}$ are determined by constitutive relations. The total dissipation is assumed to arise solely from Ohmic heating, where $\boldsymbol{J}$ is the electric current density and $\sigma$ is the electrical conductivity, since the viscous dissipation is expected to be small in planetary cores (Nimmo, 2015). Radiogenic heating contributes little entropy (Williams and Nimmo, 2004) and is not considered here.

The global energy equation for a slurry is obtained by summing the internal, mechanical and magnetic energies and integrating over the volume $V$ of the slurry. These equations are supplemented by the equations describing conservation of total mass and light element, $\xi$, which can be written (Loper and Roberts, 1977)

$$\frac{D\rho}{Dt} = -\rho\nabla \cdot \boldsymbol{u} \qquad (2)$$

and

$$\rho\frac{D\xi}{Dt} = -\nabla \cdot \boldsymbol{i} \qquad (3)$$

respectively. Here $\boldsymbol{u}$ is the fluid velocity. In the slurry the local concentration of light element depends on the local fraction of solid, $\phi$: $\xi = (1-\phi)\xi_l^s$, where $\xi_l^s$ is the concentration of light element in the liquid phase in the slurry.

Changes in the total internal energy $U$ can be expressed using the equation

$$\int \rho\frac{DU}{Dt}dV = \int \rho T\frac{Ds}{Dt}dV + \int \rho\mu\frac{D\xi}{Dt}dV + \int \frac{P}{\rho}\frac{D\rho}{Dt}dV. \qquad (4)$$

With the approximations above the total mechanical energy budget for a slurry is

$$\frac{dKE}{dt} = \int \boldsymbol{u} \cdot [\boldsymbol{F_L} - \nabla P + \rho\nabla\psi]\, dV, \qquad (5)$$

where $\psi$ is the gravitational potential, which is calculated locally as described in Davies (2015).

The total magnetic energy budget is the same as for a two-component liquid:

$$\frac{dME}{dt} = -\int \boldsymbol{u} \cdot \boldsymbol{F_L}\, dV - \int \frac{\boldsymbol{J}^2}{\sigma} dV, \tag{6}$$

where $\boldsymbol{F_L}$ is the Lorentz force. The rate of change of kinetic and magnetic energy, $dKE/dt$ and $dME/dt$ respectively, are small and can be neglected (Nimmo, 2015) from equations (5) and (6).

Adding the integral of equation (1) and (4)-(6) gives

$$\int \rho T \frac{Ds}{Dt} dV + \int \rho\mu \frac{D\xi}{Dt} dV = -\oint \boldsymbol{q} \cdot d\boldsymbol{A} + \int \rho \boldsymbol{u} \cdot \nabla\psi\, dV. \tag{7}$$

Equation (3) has been used to obtain the second term on the left-hand side and $\boldsymbol{A}$ is the (outward-pointing) area element on the surfaces that bound $V$. The first term in (7) can be rewritten using the entropy differential [equation 5.9 of Loper and Roberts (1977)],

$$ds = -\frac{\alpha}{\rho} dP + \frac{C_p}{T} dT + \bar{s} d\xi - \frac{L}{T} d\phi, \tag{8}$$

where $L$ is the latent heat, $\alpha = -1/\rho(\partial\rho/\partial T)$ is the thermal expansion coefficient and $\bar{s} = -(\partial\mu/\partial T)$ is the heat of reaction coefficient. Equation (8) is identical to the entropy differential for a binary liquid mixture except for the last term, which represents changes in entropy produced by latent heat release (absorption) when solid forms (melts).

The total energy budget for the whole core is obtained by applying equation (7) to the liquid and slurry regions and applying boundary conditions at the interface and CMB. We denote using superscripts $s$ and $l$ quantities on the snow and liquid side of the interface $r_s$ respectively. The constitutive relation for $\boldsymbol{q}$ in a binary slurry is (Loper and Roberts, 1977; Loper and Roberts, 1980)

$$\boldsymbol{q} = \mu \boldsymbol{i} + T\boldsymbol{k} = (\mu + \bar{s}T)\boldsymbol{i} - L\boldsymbol{j} - k\nabla T, \tag{9}$$

where $\boldsymbol{j}$ is the flux of solid particles and $k$ is the thermal conductivity. Note that $\boldsymbol{j} = \phi = 0$ outside the slurry. At the CMB, we assume for simplicity that there is no net mass exchange; thus

$$\boldsymbol{n} \cdot \boldsymbol{q} = -\boldsymbol{n} \cdot k\nabla T, \tag{10}$$

where the unit vector $\boldsymbol{n}$ points radially outward. To determine the boundary condition at $r_s$ we follow standard pill-box arguments (Loper and Roberts, 1987), obtaining

$$\boldsymbol{n} \cdot \langle \boldsymbol{q} \rangle = \boldsymbol{n} \cdot [\boldsymbol{q}^s - \boldsymbol{q}^l] = \rho\langle\phi\rangle[L + \xi(\mu + \bar{s}T)]\boldsymbol{n} \cdot \boldsymbol{U_s}, \tag{11}$$

where $\boldsymbol{U_s}$ is the velocity of the interface and $\langle X \rangle$ denotes the jump in the quantity $X$ across $r_s$. The terms on the right-hand side represent respectively the latent heat and heat of reaction in the shell of freezing material.

Writing $Q_{cmb} = -\oint k\nabla T \cdot \boldsymbol{n} dA$ and inserting (9)-(11) into (7) gives the global energy balance

$$Q_{cmb} = \underbrace{-\int \rho C_p \frac{DT}{Dt} dV}_{Q_S} + \underbrace{\int \alpha T \frac{DP}{Dt} dV}_{Q_P} + \underbrace{\int \rho L \frac{D\phi}{Dt} dV}_{Q_L}$$

$$\underbrace{-\int \rho(\bar{s}T + \mu)\frac{D\xi}{Dt} dV + \oint_{r_s} \rho\xi\phi[\bar{s}T + \mu]\, \boldsymbol{n} \cdot \boldsymbol{U_s} dA}_{Q_H} + \underbrace{\int \rho\boldsymbol{u} \cdot \nabla\psi\, dV}_{Q_g} + \underbrace{\oint \rho L\phi\, \boldsymbol{n} \cdot \boldsymbol{U_s} dA}_{Q_L^b}. \tag{12}$$

Here $V$ now represents the total core volume. (Note that $-\oint \boldsymbol{q} \cdot d\boldsymbol{A} = -Q_{cmb} + Q^s = -Q_{cmb} + \oint \{\boldsymbol{n} \cdot \boldsymbol{q^l} + \rho\langle\phi\rangle[L + \xi(\mu + \bar{s}T)]\boldsymbol{n} \cdot \boldsymbol{U_s}\} dA,$ and $\boldsymbol{n} \cdot \boldsymbol{q}^l$ gives the contribution to each term from the liquid region, recalling that $\phi = 0$ there.) From now on we neglect heat of reaction ($Q_H = 0$) and the small pressure changes caused by core contraction ($Q_P = 0$). The contributions to the energy and entropy balance are very small compared to the other terms (Gubbins et al., 2003; Davies, 2015).

Since the core temperature is assumed adiabatic the cooling rate at radius $r$ can be related to the CMB cooling rate (Gubbins et al., 2003)

$$\frac{DT}{Dt} = \frac{T}{T_{cmb}} \frac{dT_{cmb}}{dt}. \tag{13}$$

Assuming that the interface moves in the radial direction then $\boldsymbol{n} \cdot \boldsymbol{U_s} = -dr_s/dt$. The latent heat of melting is defined as $L = T_l \Delta s$, where $\Delta s$ is the entropy change on melting, and hence

$$Q_L^b = 4\pi r_s^2 \rho(r_s)\phi(r_s) T_l \Delta s \frac{dr_s}{dt}. \tag{14}$$

$dr_s/dt$ can be related to the core cooling rate in a manner analogous to the situation of inner core growth:

$$\frac{dr_s}{dt} = -\frac{1}{(\partial T_l/\partial P - \partial T/\partial P)} \frac{1}{\rho(r_s) g(r_s)} \frac{T}{T_{cmb}} \frac{dT_{cmb}}{dt}. \tag{15}$$

The gravitational energy $Q_g$ released due to rearrangement of light material can be re-expressed using the identity $\rho \boldsymbol{u} \cdot \nabla\psi = \nabla \cdot \rho \boldsymbol{u} \psi - \psi \nabla \cdot \rho \boldsymbol{u}$ and taking the part of the density change due to composition. We separate the contributions to $Q_g$ from the freezing out of solid in the snow zone, denoted $Q_g^s$, and remelting of solid in the liquid region, $Q_g^l$, as

$$Q_g^s = -\int \rho\psi\alpha_c \frac{\partial \xi^s}{\partial t} dV^s, \tag{16}$$

$$Q_g^l = \int \quad \rho\psi\alpha_c \frac{\partial \xi^l}{\partial t} dV^l - 4\pi r_s^2 \rho(r_s)\psi(r_s)\xi^l \frac{dr_s}{dt}, \tag{17}$$

where $\alpha_c = -1/\rho(\partial\rho/\partial\xi_l)$ is the compositional expansion coefficient for sulfur, assumed constant, and the second term on the right-hand side of (17) gives the contribution due to motion of the interface.

The sulfur concentration in the liquid region below the snow layer is obtained by applying equation (3) to the snow zone and liquid layer and adding:

$$\int \rho^s \frac{D\xi^s}{Dt} dV^s + \int \rho^l \frac{D\xi^l}{Dt} dV^l - \oint_{r_s} (\boldsymbol{i^s} - \boldsymbol{i^l}) \cdot \boldsymbol{n}\, dA = 0. \tag{18}$$

Applying a standard pill-box analysis (Loper and Roberts, 1987), the boundary condition at $r_s$ can be written

$$(\boldsymbol{i^s} - \boldsymbol{i^l}) \cdot \boldsymbol{n} = \rho(\xi^s - \xi^l)\frac{dr_s}{dt}, \tag{19}$$

where we have used the fact that the total mass of sulfur is conserved.

The time-averaging process removes the $\boldsymbol{u} \cdot \nabla$ part of the first two terms in (18). Furthermore, assuming that the liquid region is well-mixed allows $\partial\xi^l/\partial t$ to be taken outside the integral. The second term then becomes $\partial\xi^l/\partial t(\int \rho\, dV^L) = M^L \partial\xi^l/\partial t$, where $M^L$ is the mass of the liquid region. Equation (18) can then be written

$$M^L \frac{\partial \xi^l}{\partial t} = -\int \rho^s \frac{\partial \xi^s}{\partial t} dV^s + 4\pi r_s^2 \rho \left(\xi^s - \xi^l\right) \frac{dr_s}{dt}. \tag{20}$$

The second term on the right-hand of (20) is very small because $T_l$ is continuous across the interface and so $\xi^s \approx \xi^l$ ($\xi^s$ and $\xi^l$ are to be evaluated on either side of the interface). $\partial \xi^s / \partial t$ is positive because $T$, and hence $T_l$, decrease with time: more light element is needed to keep the layer at the liquidus. Therefore, as expected, $\xi^l$ decreases with time as the liquid region becomes more enriched in iron.

We obtain $\partial \xi^s / \partial t$ from the liquidus relation

$$\xi_l^s \bar{\mu} d\xi_l^s = \Delta V dP - \frac{L}{T} dT, \tag{21}$$

where $\Delta V$ is the change in volume on freezing and $\bar{\mu} = \partial \mu / \partial \xi_l^s$. $\bar{\mu}$ is calculated from ideal solution theory as $\bar{\mu} = \frac{k_b T}{\xi_l} \times Ev \times Na \times 1000/A_S$ (J kg$^{-1}$), where $k_b$ is Boltzmann's constant, $Ev$ is the electron volt, $Na$ is Avagadro's number and $A_s$ is the atomic weight of S (Gubbins et al. 2004). Solid is formed rapidly and subsequent changes in $\xi^s$ occur due to rearrangement of the solid fraction. We therefore assume that changes in $\xi^s$ occur on a timescale that is long compared to changes in $\phi$. Then $d\xi^s = (1-\phi) d\xi_l^s$ and (neglecting pressure changes)

$$\frac{\partial \xi^s}{\partial t} = -\frac{L(1-\phi)}{T \xi_l^s \bar{\mu}} \frac{\partial T}{\partial t}; \tag{22}$$

This equation resembles the relation $\frac{\partial \xi^s}{\partial t} = \frac{\partial \xi^s}{\partial T} \frac{\partial T}{\partial t}$ used by Rückriemen et al. (2015), who estimated $\frac{\partial \xi^s}{\partial T}$ from an empirical liquidus curve. Relations (20) and (22) determine $Q_g^s$ and $Q_g^l$.

On the short $\phi$ timescale we neglect variations in $\xi^s$. Then $d\phi = (1-\phi) d\xi_l^s / \xi_l^s$ and

$$\frac{D\phi}{Dt} = -\frac{L(1-\phi)}{T\left(\xi_s^l\right)^2 \bar{\mu}} \frac{DT}{Dt}. \tag{23}$$

We must distinguish between the latent heat released on freezing of solid particles, $Q_L^s$, and the latent heat absorbed on remelting, $Q_L^l$. The total mass created, $\int \rho D\phi/Dt\, dV^s$, is equal to the mass destroyed; the only difference is that freezing occurs throughout the snow zone whereas remelting occur at $r_s$. We therefore have

$$Q_L^s = \int \rho L D\phi/Dt\, dV^s \text{ , and } Q_L^l = -\int \rho L(r_s) D\phi/Dt\, dV^s.$$

Substituting equations (13)-(17) and (20)-(23) into (12) allows the global energy equation to be written symbolically as

$$Q_{cmb} = Q_s + Q_g^s + Q_g^l + Q_L^s + Q_L^l + Q_L^B = \tilde{Q} \frac{dT_{cmb}}{dt} \tag{24}$$

The additional energy sources that arise due to iron snow are the latent heat released due to formation of solid, $Q_L^s$, latent heat absorbed as falling snow remelts, $Q_L^l$, gravitational energy $Q_g^l$ released due to mixing of the remelted iron in the liquid region, and gravitational energy $Q_g^s$ released due to the sinking of iron particles in the snow zone. All terms are proportional to the CMB cooling rate as determined by equations (13), (15), (22) and (23).

The entropy equation is obtained from equation (1) in the usual way and is

$$E_J + E_k = E_s + \frac{Q_g^s + Q_g^l}{T_{cmb}} + E_L^s + E_L^l + Q_L^B \left(\frac{T_{cmb} - T(r_s)}{T_{cmb} T(r_s)}\right) = \tilde{E} \frac{dT_{cmb}}{dt} \tag{25}$$

where $E_g^l = Q_g^l/T_{cmb}$, $E_g^s = Q_g^s/T_{cmb}$, and

$$E_J = \int \frac{\boldsymbol{J} \cdot \boldsymbol{J}}{\sigma T} dV,$$

$$E_k = \int k \left(\frac{\nabla T}{T}\right)^2 dV,$$

$$E_s = -\int \rho C_p \left(\frac{1}{T_{cmb}} - \frac{1}{T}\right) \frac{DT}{Dt} dV$$

$$E_L^s = \int \rho L \frac{D\phi}{Dt} \left(\frac{1}{T_{cmb}} - \frac{1}{T}\right) dV^s,$$

$$E_L^l = -\int \rho L(r_s) \frac{D\phi}{Dt} \left(\frac{1}{T_{cmb}} - \frac{1}{T}\right) dV^s.$$

Viscous and chemical dissipations are thought to be much smaller than $E_J$ and so are neglected (Nimmo 2015). Equations (24) and (25) are evolved forward in time using a timestep of 1 Myr. The initial time is 4.5 Ga and the final time is the present-day unless a snow zone occupies the whole core in which case the calculation is terminated at that point. The thermal and chemical evolution of the coupled snow-liquid system is calculated such that the base of the snow zone $r_s$ is at the liquidus temperature at each time step. This evolution repeats at each model iteration as the core cools.

Model Parameters

Mantle convection sets $Q_{cmb}$ while core convection sets the CMB temperature and so the evolution of the two systems should strictly be solved simultaneously. However, significant uncertainties in the parameterization of mantle convection, particularly the appropriate rheological law and the scaling of surface and CMB heat flow with temperature, mean that we do not expect to obtain a definitive thermal history for Mars but seek to understand whether the snow regime is potentially compatible with existing geodetic and magnetic observations. Focusing on the core alone allows us to elucidate the individual effects of the various physical processes that arise from snow zone growth.

We consider two time-series of $Q_{cmb}$ from previous studies that both match the inferred dynamo cessation time, but nevertheless exhibit significant differences that embody some of the uncertainties in the mantle problem. The time-series of Williams and Nimmo (2004) ( hereafter W04) is from a parameterized model of stagnant lid mantle convection, while L14 (Leone et al. 2014) was calculated from a 3D thermochemical mantle convection simulation. To enable a flexible implementation we approximate the time-dependence of the W04 and L14 $Q_{cmb}$ time-series by three piecewise linear segments that represent the initial rapid decline, the near-constant variation in recent times, and the intermediate transition period (Figure 2b).

We vary core properties individually to elucidate their influence on the snow regime. This has the potential to produce an inconsistency since both W04 and L14 used a particular core model, which produced a particular time-series of $T_{cmb}$ that is compatible with their time-series of $Q_{cmb}$. To mitigate against this effect we set the initial CMB temperature, $T_{init}$, to be close the original values. For W04 $T_{init} = 2400$ K, while L14 did not quote a value and so we vary $T_{init}$ around the W04 value. We find that deviations in $T_{cmb}$ after 4.5 billion years of evolution differ by $< 50$ K from the original values in the majority of models. The paucity of independent observational constraints leads to some interdependencies between estimates of interior structure properties (e.g. assuming a temperature profile in order to estimate the density profile), but in this initial exploration we vary each parameter independently.

Values of density $\rho$, CMB radius $r_{cmb}$ and CMB pressure $P_{cmb}$ (Table A1) are selected from W04 and also from a recent detailed analysis of the Martian interior (Rivoldini et al. 2011) that produced a range of models constrained by moment-of-inertia and $k_2$ Love number data. Rivoldini *et al.* (2011) find that $\rho$ varies by only $5 - 10\%$ across the Martian core and so there is little error in taking $\rho$ constant. These values determine the structure of the Martian core, i.e. the radial profiles of gravity, gravitational potential and hydrostatic pressure (Figure 2). For each model the pressure scale constructed in this manner is used to establish the temperature profiles discussed below.

The Martian core is thought to be composed primarily of an iron-sulfur alloy (Dreibus and Wanke 1985) and this simple chemistry has been used in almost all thermal history models to date (Breuer and Moore 2015). A more complicated core chemistry might be expected since the high temperatures achieved during the early stage of core formation may have facilitated the incorporation of Si, O, Ni, P, O, and H into liquid iron (Tsuno et al. 2007). The Martian core is expected to contain a negligible amount of Si (Sanloup and Fei, 2004) and O (Tsuno et al. 2007; Rubie et al. 2011), while Ni has only a minor effect on phase equilibria of Fe-S (Stewart et al. 2007). The amount of phosphorus in the Martian core is thought to be ten times that of Earth's core (0.16 vs. 0.02 wt% $P_2O_5$) (Dreibus and Wanke, 1985; Hart and Zindler, 1986). Both its abundance and the magnetic transitions of P-bearing phases may influence density distribution in the core (Gu et al., 2014), but we do not consider this additional complexity here. The density and melting point may also be lowered by the presence of hydrogen, though its content in the core is poorly constrained. In the absence of sufficient constraints regarding core equilibrium chemistry or a suitable theory for the melting point depression in ternary (or higher order) mixtures we model the evolution of a Fe-S mixture. The initial sulfur concentration is varied between 10 and 15%, which is within the estimates of previous models of the composition of Mars (e.g. Dreibus and Wanke, 1985; Taylor, 2013).

The adiabatic temperature is parameterized by the equation $T = T_0(1 + 0.02P)$, which fits the published profiles of Williams and Nimmo (2004) and Fei and Bertka (2005). The coefficient $T_0$ varies as the core cools. Note that the adiabatic gradient $\partial T/\partial r$ is proportional to $T$ and therefore decreases as the core cools (Figure 2c).

The liquidus temperature $T_l$ is parameterized following Williams and Nimmo (2004)

$$T_l = T_{l0}(1 - \xi + \xi_0)(1 + T_{l1}P + T_{l2}P^2) \tag{26}$$

where $\xi_0$ is the initial sulfur concentration (i.e. the concentration at $t = 4.5$ Ga) and pressure is measured in GPa. The form (26) was chosen to allow comparison with previous results and to benchmark the code. The constants $T_{l0-2}$ are obtained from least-squares fits (setting $\xi = \xi_0$) to experimental results performed on samples containing 10.6 wt% S (Stewart *et al.*, 2007) and 14.2 wt% S (Stewart et al., 2007) (Figure 2d), which are denoted S07-10.6 and S07-14.2 respectively.

The latent heat $L$ released on freezing is $L = T_l \Delta s$. The entropy of melting, $\Delta s$, is parameterized by the equation (Davies, 2015) $\Delta s = 1.99731 - 0.0082P$ where the constant coefficients are obtained by a least-squares fit to the data of Figure 3 in Alfè et al. (2002) with the free energy correction applied. Only data in the range $50 - 70$ GPa, the lowest values considered by Alfè et al. (2002), were used in the fitting since $P$ as the center of Mars is about 40 GPa.

## 3. Results

Input parameters for all models are listed in Table A1 and diagnostics are presented in Table A2. We focus first on a model that uses the default parameters of W04 except for the melting curve and initial S concentration, which are set using the S07-10.6 profile. In this model the dynamo cessation time $D_f = 459$ Myrs (~4 Ga), defined by $E_J$ falling below zero, while the present-day CMB temperature $T_{cmb}^{pres} = 1822$ K (Table A2, highlighted in bold); both values are very close to the original solution obtained by W04. Figure 3 shows profiles of temperature, solid fraction and S concentration for this model. Approximately 3.2 billion years into the evolution, $T_{cmb}$ falls below $T_l$ at the CMB and an iron snow layer begins to form and grows to $146$ km by the present day. The solid fraction $\phi$ remains below ≈ 0.2 %, consistent with the modeling assumptions and with a recent model of iron snow in Ganymede's core (Rückriemen, et al., 2015), though the profiles of $\phi$ do not exhibit the curvature obtained by Rückriemen et al., (2015) near the base of the layer, which appears to stem from the different methods used to estimate $\partial\xi^s/\partial T$. The sulfur concentration increases across the snow zone by almost a factor of 1.5 at the present day, which arises partly due to the decreased S concentration in the deep core as Fe remelts and partly because of the enrichment in S with radius required to keep the snow zone on the liquidus.

Figure 4 shows the contributions of individual terms to the energy and entropy balances for the model in Figure 3. The latent heat terms $Q_l^s$ and $Q_L^l$ make an order of magnitude larger contribution to the energy budget than the gravitational energy terms $Q_g^s$ and $Q_g^l$ in agreement with the study of Rückriemen et al. (2015), while $Q_l^b$ is negligible. $Q_L^s$ and $Q_L^l$ almost balance since the same amount of mass is produced and destroyed; the small difference arises because the latent heat coefficient varies with depth. Therefore, these terms have little impact on the cooling rate at the onset of snow formation. The smallness of $Q_g^s$ and $Q_g^l$ reflects the slowness of compositional changes because the cooling rate is low and $\xi^l$ is a weak function of $T$ at conditions corresponding to the upper region of the Martian core.

The high thermodynamic efficiency of $Q_g^s$ and $Q_g^l$ means that the corresponding entropies are comparable to $E_L^s$ and $E_L^l$. Nevertheless, the overall entropy produced from the formation and remelting of iron snow is small (Figure 4) and the dynamo does not restart as long as the snow zone remains relatively thin. The dynamo only restarts when the entropy produced by gravitational energy release due to the remelting snow, $Q_g^l$, which is proportional to the snow zone volume and growth rate [equations (17) and (20)], overcomes the conduction entropy $E_k$. Rückriemen et al. (2015) inferred that dynamo action arose in their models of Ganymede. This finding is not comparable to our results since they used scaling laws to assess the onset and maintenance of dynamo action, rather than the entropy formulation employed here. Table A2 shows that the dynamo restarts when $r_{cmb} - r_s > 400$ km in our suite of models.

Figure 5 shows a solution obtained with the same parameters as the model in Figure 3, except with a lower value of $\rho$ (highlighted in italics in Table A2). Lowering $\rho$ reduces $\widetilde{Q_s}$ [equation (24)] and therefore leads to faster cooling at early times and an older snow zone. The effect of decreasing $\rho$ from 7000 kg m$^{-3}$ to 6000 kg m$^{-3}$ is significant, which might partly reflect the lack of feedback on $Q_{cmb}$ due to changes in $T_{cmb}$ in our model; however, decreasing $\rho$ also decreases the difference in gravity, gravitational potential, pressure, and adiabatic temperature across the core and so affects all terms in the energy and entropy balances. The continual enrichment of the upper layer and continual depletion of the lower layer in light element leads to rapid growth of the snow zone once it reaches a critical depth. In this final stage the latent heat terms increase rapidly as both are proportional to $(\xi_l^s)^{-2} \approx \left(\xi^l\right)^{-2}$ near $r_s$ [equation (23)], while the increase in gravitational energy is less dramatic since $\partial \xi^s / \partial t$ is independent of $\xi$ [equation (22)] and the effect of decreasing liquid mass is partly compensated by increasing snow zone mass. The gravitational energy released by rapid variation in sulfur concentration provides sufficient entropy to restart the dynamo; this solution is not in agreement with the existing constraints on Mars' thermal history.

The constraint that the Martian dynamo cannot restart (negative $E_J$ at the present-day) places a nominal upper bound on the thickness of the present-day snow zone of $\sim 400$ km based on the limited model set available (Table A2). Occasionally, models with thick snow zones can produce thin layers below the CMB where $\xi$ exceeds the eutectic composition of $\approx 16$ wt % at $P \approx 20$ GPa (Stewart *et al.* , 2007). The dynamics of this scenario are not included in our model, but it would produce light solid that floats to the CMB, thus reducing the estimates of gravitational energy compared to our calculations. However, the fact that such layers are very thin suggests that the associated entropy reduction would not prevent the dynamo from restarting.

Figure 6 demonstrates the influence of parameter variations on the snow layer for models. Here the label for each symbol denotes the single quantity that was changed compared to the default model, which used the parameters highlighted in Table A1. The inferred dynamo cessation time ($D_f$) is relatively insensitive to changes in $T_{init}$, $\rho$, $P_{cmb}$ and $r_{cmb}$, but is very sensitive to changes in $k$; in Figure 6 the dynamo fails too late with $k = 30$ W m$^{-1}$ K$^{-1}$ and too early with $k > 50$ W m$^{-1}$ K$^{-1}$. Aside from these cases all models in Figure 6 match the inferred dynamo cessation time, produce present-day CMB temperatures well above the eutectic value (Table A2) of $1300 - 1500$ K at Martian CMB pressures (Rivoldini et al., 2011), and produce thin snow zones consistent with geodetic observations that suggest a predominantly liquid present-day core (Yoder et al., 2003). The iron snow regime is less likely to emerge for larger $r_{cmb}$ and certain $Q_{cmb}$ time-series, which both cause the core cooling rate to decrease, though we obtained snow zones with all $r_{cmb}$ and $Q_{cmb}$ values tested. Our results predict a strong sensitivity to $T_{init}$; however, this may be an artifact of the model assumption that $Q_{cmb}$ does not change with $T_{cmb}$. The crucial parameter is the initial S concentration $\xi_0$, which determines the melting temperature and therefore strongly influences the initial difference between adiabatic and liquidus temperatures at the CMB

Finally we consider whether iron snow zones arise using the preferred interior structure model of Rivoldini *et al.* (2011) and other default parameter values in Table A1. These runs use $\xi_0 = 0.142$ and the S07-14.2 melting curve (Table A2). The high values of $\xi_0$ and $r_{cmb}$ do not favor iron snow formation, but we do find relatively thin present-day snow zones in models with $T_{init} \approx 2250$ K, approximately 150 K below the value used in W04. This value still suggests a core that was initially superheated with respect to the mantle, consistent with the original modeling assumptions, and we have not attempted to 'optimize' our solution through a systematic parameter search as the uncertainties in several key variables do not warrant such a procedure. This model has a relatively late dynamo cessation time of 3.6 Ga, but increasing $k$ to 50 W m$^{-1}$ K$^{-1}$, which is well within uncertainty, provides an acceptable value of 4 Ga while leaving the snow zone evolution unchanged.

## 4. Application of the snow model to the Martian core

### *4.1. Fast melting and remelting*

Relaxing the fast melting approximation (i.e. incorporating departures from phase equilibrium) introduces additional terms and equations into the slurry theory and drastically increases the complexity of the constitutive relations (Loper and Roberts, 1977). These additional effects require macroscopic parameterizations of microscopic processes (Loper, 1992) that are poorly understood and the resulting terms are hard to estimate for planetary cores. While the overall influence of fast melting is hard to quantify, we might expect that the effects may not be significant as long as the relaxation to phase equilibrium occurs on timescales that are much shorter than the long timescale of interest (Gyrs). Incorporating the effects of a multi-component solid phase also significantly complicates the theory by requiring that the history of individual particles is modeled. At present we believe both approximations are sensible compromises for modeling the long-term behavior of snow layers in planetary interiors.

Rückriemen et al (2015) used scaling laws with simple assumed particle sizes and geometries to argue that the fast remelting approximation is appropriate for modeling iron snow in Ganymede's core. Solomatov and Stevenson (1993) analyzed the conditions required to perpetually suspend particles in a magma ocean, but our model does not predict quantities such as the convective velocity needed to apply their theory. If some of the solid particles remain suspended in the snow zone on long timescales the latent heat released on freezing, $Q_L^s$, will exceed the latent heat absorbed on remelting, $Q_L^l$. Since $Q_L^s$ is released close to the CMB it has a low thermodynamic efficiency factor, suggesting a reduction in entropy available to power the dynamo compared to the fast remelting case considered in this paper. It therefore appears that relaxing the fast remelting assumption would not significantly change our results, though hydrodynamic simulations of slurry dynamics are needed to test the veracity of this statement.

### *4.2. Thermal stratification at the CMB*

The demise of the Martian dynamo is signified by the Ohmic dissipation $E_J$ dropping below zero. However, $E_J \geq 0$ by definition and so negative values indicate an inconsistency in the modeling assumptions. The fact that $Q_{cmb} < Q_a$ for most of the evolution suggests that thermal stratification ensues and the temperature profile deviates from the assumed adiabat profile near the top of the core in order to balance the entropy budget with $E_J = 0$ after the dynamo fails. If $E_J = 0$ prior to snow zone formation, the gravitational entropy terms $E_g^s, E_g^l > 0$ (Figures 4 and 5) that arise during snow zone growth would make $E_J > 0$ and potentially restart the dynamo. Strictly, $E_J$ must exceed some minimum value, denoted $E_J^m$, for dynamo action to occur. $E_J^m$ is hard to estimate because it depends on the magnetic field morphology in the core, including the small-scale fields and the field components that remain inside the core, neither of which can be observed. Using just the observable field at the CMB gives $E_J^m \sim 10^6$ MW/K (Gubbins, 1975; Backus et al., 1996), similar to the values of $E_g^s$ and $E_g^l$ in our models (Figures 4 and 5). The real value of $E_J^m$ is likely to be higher than this (Nimmo, 2015), suggesting that snow zone growth would not restart the dynamo. Since the dominant contributions to $E_J$ come from small-scale magnetic fields inside the core it may be possible that some field generation accompanies snow zone formation but produces an extremely weak signal at the planet's surface.

As discussed in Section 2, the thermally stable layer receives more heat through its base than can be removed at the CMB. Thus, the layer must heat up and the CMB temperature should be higher than predicted by our model, raising the question of whether the snow zone would still form. To address this issue we must first determine the relevant equations governing temperature variations in the conducting region. The temperature equation in a slurry ignoring pressure, radiogenic, and dissipative effects and assuming constant material properties is (Loper and Roberts, 1987)

$$\rho C_p \frac{DT}{Dt} = k\nabla^2 T + L\nabla \cdot \boldsymbol{j} + \rho L \frac{D\phi}{Dt} \tag{27}$$

where $\boldsymbol{j} < 0$ is the downward flux of solid material. The last two terms represent the total rate of change of solid mass per unit volume. A detailed analysis is complicated because $\boldsymbol{j}$ depends on the size and distribution of solid particles. However, we observe that the fast melting approximation requires that solid freezes out quickly ($\frac{D\phi}{Dt} > 0$) while fast remelting requires that solid falls out of the layer quickly ($\nabla \cdot \boldsymbol{j} < 0$) compared to the long timescale over which the temperature is changing. Since all solid leaves the layer after each timestep, on this long timescale the last two terms are expected to cancel out, leaving a standard diffusion equation for the temperature.

To estimate the temperature difference between an adiabatic and thermally stratified region , we first consider an infinite half-space with prescribed time-independent subadiabatic heat-flux at the boundary and zero initial temperature (corresponding to no departure from an initial adiabatic profile). In this case, the solution to (27) without solid ($\phi = \boldsymbol{j} = 0$) gives a boundary temperature $T_0$ that can be written

$$T_0(t) = 2\frac{(Q_a - Q_{cmb})}{4\pi r_c^2 k}\sqrt{\frac{\kappa t}{\pi}} \tag{27}$$

(Carslaw and Jaeger, 1959) where $\kappa = k/\rho C_p$ is the thermal diffusivity. In Figure 4, a thermally stable layer starts to grow at time $t_t = 400$ Myrs into the calculation and a snow zone formed at approximately $t_s = 3.2$ billion years, giving $t = t_s - t_t = 2.8$ Gyrs. At time $t_s$, $Q_{cmb} = 0.257$ TW and $Q_a = 0.875$ TW, corresponding to the strongest subadiabatic conditions (Figure 4). With these values equation (27) shows that thermal conduction increases the CMB temperature by $\Delta T = T_0(t_s) - T_0(t_t) \approx 100$K over 2.8 Gyrs above the value predicted from cooling on an adiabat. Using values for other models that match the cessation time for the Martian dynamo inferred from magnetic observations (Table A2) gives $\Delta T = 100 - 250$ K.

The analytical expression (27) ignores the effects of spherical geometry, finite stable layer thickness and temporal changes in CMB heat flow. We account for these effects by numerically solving the 1D conduction equation $\partial T/\partial t = \kappa r^{-2} d/dr(r^2 dT/dr)$ in a spherical shell of thickness $L$. Using the time-series of $Q_{cmb}$ in Figure 4 and an initial adiabatic $T$ from this run at time $t_t$ we obtain $\Delta T = 70 - 140$ K for $100 \leq L \leq 700$ km. The cooling rate in our models is 70-150 K Gyrs$^{-1}$ at the time of snow zone formation and the snow zones form 0.5-1.5 Gyrs before present, suggesting that snow zone formation would be delayed until the recent past. However, all of these $\Delta T$ values are over-estimates because they ignore movement of the stable layer interface and omit the reduction in core cooling rate induced by stratification and by entrainment due to the underlying convection. We conclude that thermal stratification could delay, but not prevent, the onset of snow formation, though a more complete model of these effects is clearly needed.

The adiabatic temperature profile used in our calculations ignores the effect of a stabilizing compositional gradient, the second term in the relation $\frac{dT}{dr} = -\frac{\alpha g T}{C_p} - \frac{T\bar{s}}{C_p}\frac{d\xi}{dr}$. The first term, calculated directly from the models, is $\approx\ 0.6 - 1$ K km$^{-1}$ at the CMB. Using ideal solution theory gives $\bar{s}\ = -k_B\, log(\bar{\xi}) \times\ Ev \times\ Na \times \frac{1000}{32} \approx\ 260\, J\, K^{-1}\, kg^{-1}$ assuming a molar sulfur fraction $\bar{\xi} = 0.1$. Taking $T = 2000$ K, $C_p = 780$ J K$^{-1}$kg$^{-1}$ (Table A1) and $d\xi/dr \approx 4 \times 10^{-7}$m$^{-1}$ from Figure 2 means that the second term is $\approx$ 0.2 K km$^{-1}$. This is an overestimate since $d\xi/dr$ depends on $dT/dr$ in the model as discussed above, suggesting that the 'dry' adiabat assumed in the modelling is a good approximation to the 'wet' adiabat that includes compositional variations.

Departures from an adiabatic temperature profile affect the energy budget mainly through the $Q_s$ term since this involves an integral over $T$. To quantify the effect, we consider for simplicity a linear subadiabatic profile in the top 100 km of the present-day core with the CMB temperature 140 K above an adiabat, corresponding to the most extreme estimates above. The resulting 5% decrease in $Q_s$ produces a change in the cooling rate of ~1 K Gyr$^{-1}$. Changes in sulfur concentration and solid fraction will also decrease in the presence of thermal stratification as the terms are proportional to $T^{-1}$ (equations (22) and (23)), but the overall effect on core cooling rate, and hence snow zone growth rate, is very small compared to other uncertainties in the calculation. The $E_s$ term in the entropy balance is reduced by a greater amount that $Q_s$, but this effect is countered by a reduction in $E_k$ since the conduction profile is shallower and hotter than an adiabat, resulting in a small change to the predicted dynamo entropy. Even weaker effects are predicted at earlier times or for younger stable layers. These simple calculations suggest that the assumption of neglecting departures from the adiabat in the energy-entropy balance is justified.

Finally, we expect that the presence of thermal stratification would reduce our estimates of the gravitational energy $Q_g^s$ generated by migration of solid within the snow zone, though our calculations suggest that this term makes a negligible contribution to the overall energy and entropy budgets when the snow zone is only a few hundred km (Figure 4). Nevertheless, the current parameterization of $Q_g^s$ is simple at best and would benefit greatly from new experimental and/or numerical studies.

## 5. Conclusions

The presence of an approximately 100-km-thick snow layer at the top of the Martian core is consistent with the planets' magnetic history and available observational constraints on its core structure, temperature and composition. Snow zone nucleation is favored for lower initial sulfur concentrations and core temperatures and for smaller core sizes. Snow zones that grow to $\approx$ 400 km produce enough gravitational energy to restart the dynamo, suggesting that this is an upper limit on the layer depth in the Martian core.

Future work simulating slurry dynamics with and without thermal stratification should enable improved parameterizations of the thermal and compositional profiles in the snow zone and the gravitational energy terms in the energy balance and therefore mitigate the relaxation of some assumptions invoked in this study. Future parameterized models could also include coupled core-mantle evolution. Considering the core in isolation has allowed us to focus on snow zone dynamics, divorced from the complexities and uncertainties in mantle evolution modeling, but at the expense of being restricted to a narrow range of CMB heat flow time-series and initial core temperatures. In particular, if solutions satisfying the available constraints can be obtained with lower initial CMB temperatures it will be possible to obtain thicker present-day snow zones than we have found.

Snow layers would not support seismic shear waves owing to the spatially dispersed nature of the solid phase, but could affect the core density. If these differences can be detected by observations from future spacecraft missions, it will provide profound inside into the thermochemical evolution of the Martian interior.

**Acknowledgments:** CJD is supported by a Natural Environment Research Council Independent Research Fellowship (NE/L011328/1) and a Green Scholarship at UC San Diego-SIO-IGPP. The authors are grateful to Cathy Constable, Jon Mound and the Deep Earth Research Group at Leeds for many constructive comments, Sam Greenwood for calculating the conduction solution for a thermally stable layer, and two reviewers whose detailed and thoughtful comments helped improve the manuscript.

**Table A1**. Input parameters used in the thermal history model. Gravity $g$, gravitational potential $\psi$ and pressure $P$ are derived from the density assuming hydrostatic balance. Density is assumed depth-independent as interior structure models predict only $5 - 10\%$ variation across the Martian core (Rivoldini et al. 2011); the constant value $7211$ kg m$^{-3}$ was sometimes used instead of the Williams and Nimmo value of $7011$ kg m$^{-3}$ as this accounts for the increase of $\rho$ with depth in their model. The thermal expansion coefficient $\alpha$, heat of reaction coefficient $\bar{s}$ and volume change on melting $\Delta V$ that appear in the governing equations are not included because the terms in the energy balance in which they appear are small enough to neglect. The latent heat is $L = T_l \Delta s$. Bold indicates the default value when multiple values have been used. Here W04 refers to Williams and Nimmo (2004), F05 is Fei and Bertka (2005) and S07 is Stewart *et al.* (2007).

| Quantity | Symbol | Units | Value | Reference |
|---|---|---|---|---|
| Density | $\rho$ | kg $m^3$ | 7011 (**7211**) | W04 |
| | | | 6000-6500 | (Rivoldini et al. 2011) |
| CMB radius | $r_{cmb}$ | km | **1627** | W04 |
| | | | $1794 \pm 65$ | (Rivoldini et al. 2011) |
| CMB pressure | $P_{cmb}$ | GPa | **21** | W04 |
| | | | $19 - 23$ | (Rivoldini et al. 2011) |
| Entropy of melting | $\Delta s$ | $k_B$ | $\Delta s = 1.99731 - 0.0082P$ | (Alfè et al. 2002) |
| Adiabatic temperature | $T$ | K | $T = T_0(1 + 0.02P)$ | W04, F05 |
| Thermal conductivity | $k$ | W $\mathrm{m}^{-1}\ \mathrm{K}^{-1}$ | $20, \mathbf{40}, 60$ | W04; Deng *et al.*, (2013) |
| Specific Heat Capacity | $C_p$ | J $\mathrm{kg}^{-1}\ \mathrm{K}^{-1}$ | 780 | W04 |
| Compositional expansion coefficient | $\alpha_c$ | - | 0.64 | (Gubbins et al. 2004) |
| S concentration | $\xi$ | - | See text | S07 |
| Liquidus temperature | $T_l$ | K | $\mathbf{T_{l0} = 1990.5, T_{l1} = -0.0022, T_{l2} = 3.8}$**e-7 S07 (10.6% S)**<br>$T_{l0} = 1860.2, T_{l1} = -0.00512, T_{l2} = -1.226$e-5 S07 (14.2% S) | |

**Table A2**. Summary of models conducted for this study. Density $\rho$ (kg m$^{-3}$), thermal conductivity (W m$^{-1}$ K$^{-1}$), pressure at the core-mantle boundary (CMB) $P_{cmb}$ (GPa), CMB radius $r_{cmb}$ (km), initial temperature at 4.5 Ga $T_{init}$ (K), liquidus profile $T_l$ and CMB heat flow time-series $Q_{cmb}$ are model inputs. The model outputs are the time at which the dynamo failed $D_f$ (Myrs), the time $t_t$ (Myrs) corresponding to onset of thermal stratification below the CMB, the time $t_s$ (Myrs) corresponding to snow zone nucleation ($t_s = 0$ implies no snow zone formed), present-day radius of the snow zone $r_s$ (km), present-day CMB temperature $T_{cmb}^{pres}$ (K), present-day S concentration at the CMB $\xi^{pres}$ (K) and present-day Ohmic dissipation (MW K$^{-1}$). Runs highlighted in red match the estimated time of $4.1 - 3.8$ Ga ($D_f = 400 - 700$ Myrs) for termination of the dynamo; bold indicates runs shown in Figures 3-4 and italics (row 1) indicates the model shown in Figure 5.

| $\rho$ | $k$ | $P_{cmb}$ | $r_{cmb}$ | $T_{init}$ | $T_l$ profile | $Q_{cmb}$ | $D_f$ | $t_t$ | $t_s$ | $r_s$ | $T_{cmb}^{pres}$ | $\xi^{pres}$ | $E_J^{pres}$ |
|---|---|---|---|---|---|---|---|---|---|---|---|---|---|
| ***6000*** | ***40*** | ***21*** | ***1627*** | ***2400*** | ***0.106_SSW07*** | ***WN04*** | ***1112*** | ***1076*** | ***2056*** | ***0*** | ***N/A*** | ***N/A*** | ***N/A*** |
| 6200 | 20 | 19 | 1800 | 2400 | 0.142_SSW07 | WN04 | 2390 | 2214 | 0 | 1800 | 1786 | 0.14 | -8 |
| 6200 | 40 | 19 | 1800 | 2400 | 0.142_SSW07 | LT13 | 792 | 760 | 0 | 1800 | 1950 | 0.14 | -30 |
| 6200 | 40 | 19 | 1800 | 2200 | 0.142_SSW07 | WN04 | 940 | 882 | 3159 | 1637 | 1587 | 0.19 | -16 |
| 6200 | 40 | 19 | 1800 | 2250 | 0.142_SSW07 | WN04 | 896 | 837 | 3870 | 1741 | 1636 | 0.16 | -24 |
| 6200 | 40 | 19 | 1800 | 2300 | 0.142_SSW07 | WN04 | 850 | 788 | 0 | 1800 | 1686 | 0.14 | -25 |
| 6200 | 40 | 19 | 1800 | 2400 | 0.142_SSW07 | WN04 | 760 | 693 | 0 | 1800 | 1786 | 0.14 | -26 |
| 6200 | 50 | 19 | 1800 | 2250 | 0.142_SSW07 | WN04 | 482 | 464 | 3870 | 1741 | 1636 | 0.16 | -33 |
| 6200 | 40 | 21 | 1627 | 2400 | 0.142_SSW07 | WN04 | 1004 | 958 | 0 | 1627 | 1708 | 0.14 | -14 |
| 6200 | 40 | 21 | 1800 | 2400 | 0.106_SSW07 | WN04 | 810 | 747 | 2804 | 1518 | 1789 | 0.16 | -8 |
| 6200 | 40 | 21 | 1800 | 2300 | 0.142_SSW07 | WN04 | 896 | 837 | 0 | 1800 | 1684 | 0.14 | -24 |
| 6200 | 40 | 21 | 1800 | 2400 | 0.142_SSW07 | WN04 | 810 | 747 | 0 | 1800 | 1784 | 0.14 | -24 |
| 6500 | 40 | 21 | 1627 | 2400 | 0.106_SSW07 | WN04 | 832 | 774 | 2470 | 1138 | 1758 | 0.18 | 11 |
| 7000 | 40 | 21 | 1627 | 2400 | 0.106_SSW07 | WN04 | 508 | 486 | 2943 | 1425 | 1802 | 0.16 | -16 |
| 7100 | 40 | 21 | 1800 | 2400 | 0.106_SSW07 | WN04 | 423 | 396 | 3996 | 1767 | 1876 | 0.12 | -44 |
| 7211 | 40 | 18 | 1627 | 2400 | 0.106_SSW07 | WN04 | 446 | 423 | 3020 | 1471 | 1824 | 0.15 | -24 |
| 7211 | 40 | 19 | 1627 | 2400 | 0.106_SSW07 | WN04 | 454 | 432 | 3069 | 1476 | 1823 | 0.15 | -24 |
| **7211** | **40** | **20** | **1627** | **2400** | **0.106_SSW07** | **WN04** | **459** | **436** | **3118** | **1481** | **1822** | **0.15** | **-24** |
| 7211 | 20 | 21 | 1627 | 2400 | 0.106_SSW07 | WN04 | 1467 | 1431 | 3168 | 1487 | 1821 | 0.15 | -5 |
| 7211 | 30 | 21 | 1627 | 2400 | 0.106_SSW07 | WN04 | 954 | 896 | 3168 | 1487 | 1821 | 0.15 | -14 |
| 7211 | 40 | 21 | 1627 | 2300 | 0.106_SSW07 | LT13 | 666 | 630 | 3622 | 1590 | 1875 | 0.12 | -31 |
| 7211 | 40 | 21 | 1627 | 2400 | 0.106_SSW07 | LT13 | 616 | 576 | 0 | 1627 | 1975 | 0.11 | -32 |
| 7211 | 40 | 21 | 1627 | 2300 | 0.106_SSW07 | WN04 | 495 | 477 | 2016 | 1196 | 1737 | 0.19 | 1 |
| 7211 | 40 | 21 | 1627 | 2350 | 0.106_SSW07 | WN04 | 482 | 459 | 2552 | 1364 | 1776 | 0.17 | -14 |
| 7211 | 40 | 21 | 1627 | 2400 | 0.106_SSW07 | WN04 | 468 | 446 | 3168 | 1487 | 1821 | 0.15 | -23 |
| 7211 | 40 | 21 | 1627 | 2450 | 0.106_SSW07 | WN04 | 454 | 432 | 3924 | 1581 | 1869 | 0.12 | -28 |
| 7211 | 40 | 21 | 1627 | 2500 | 0.106_SSW07 | WN04 | 441 | 418 | 0 | 1627 | 1919 | 0.11 | -29 |
| 7211 | 40 | 21 | 1730 | 2400 | 0.106_SSW07 | WN04 | 428 | 405 | 3726 | 1673 | 1861 | 0.13 | -37 |
| 7211 | 40 | 21 | 1800 | 2300 | 0.106_SSW07 | LT13 | 544 | 495 | 0 | 1800 | 1923 | 0.11 | -52 |
| 7211 | 40 | 21 | 1800 | 2350 | 0.106_SSW07 | LT13 | 518 | 464 | 0 | 1800 | 1973 | 0.11 | -52 |
| 7211 | 40 | 21 | 1800 | 2400 | 0.106_SSW07 | LT13 | 490 | 436 | 0 | 1800 | 2023 | 0.11 | -52 |
| 7211 | 40 | 21 | 1800 | 2450 | 0.106_SSW07 | LT13 | 459 | 405 | 0 | 1800 | 2073 | 0.11 | -52 |
| 7211 | 40 | 21 | 1800 | 2400 | 0.106_SSW07 | WN04 | 405 | 374 | 4190 | 1782 | 1886 | 0.11 | -47 |
| 7211 | 50 | 21 | 1627 | 2400 | 0.106_SSW07 | WN04 | 315 | 284 | 3168 | 1487 | 1821 | 0.15 | -32 |
| 7211 | 60 | 21 | 1627 | 2400 | 0.106_SSW07 | WN04 | 148 | 108 | 3168 | 1487 | 1821 | 0.15 | -41 |
| 7211 | 40 | 22 | 1627 | 2400 | 0.106_SSW07 | WN04 | 477 | 454 | 3218 | 1492 | 1820 | 0.15 | -23 |
| 7211 | 40 | 23 | 1627 | 2400 | 0.106_SSW07 | WN04 | 486 | 464 | 3267 | 1498 | 1819 | 0.14 | -23 |
| 7400 | 40 | 21 | 1627 | 2400 | 0.106_SSW07 | WN04 | 436 | 414 | 3384 | 1529 | 1838 | 0.14 | -29 |
| 7500 | 40 | 21 | 1627 | 2400 | 0.106_SSW07 | WN04 | 418 | 392 | 3506 | 1547 | 1846 | 0.13 | -31 |

**Figure 1**. (**A**) Heat and entropy sources used to calculate the evolution of the Martian core and dynamo. $Q_S$ is the secular cooling. Top-down crystallization leads to latent heat release as solid iron forms throughout the snow zone, $Q_L^s$, latent heat absorption $Q_L^l$ as iron snow remelts at the top of the liquid region and gravitational energy release due to the negative buoyancy of iron sinking in the snow zone ($Q_g^s$) and remelting at the top of the liquid region ($Q_g^l$). Each heat source has an associated entropy term. The entropy balance contains additional contributions from thermal conduction ($E_k$) and Ohmic dissipation ($E_J$); the latter determines the viability of dynamo action. **(B)** Example model run showing growth of a snow zone. The adiabatic temperature and melting curve (Stewart et al. 2007) both evolve with time and the radius where they intersect defines the instantaneous base of the snow zone.

**Figure 2**

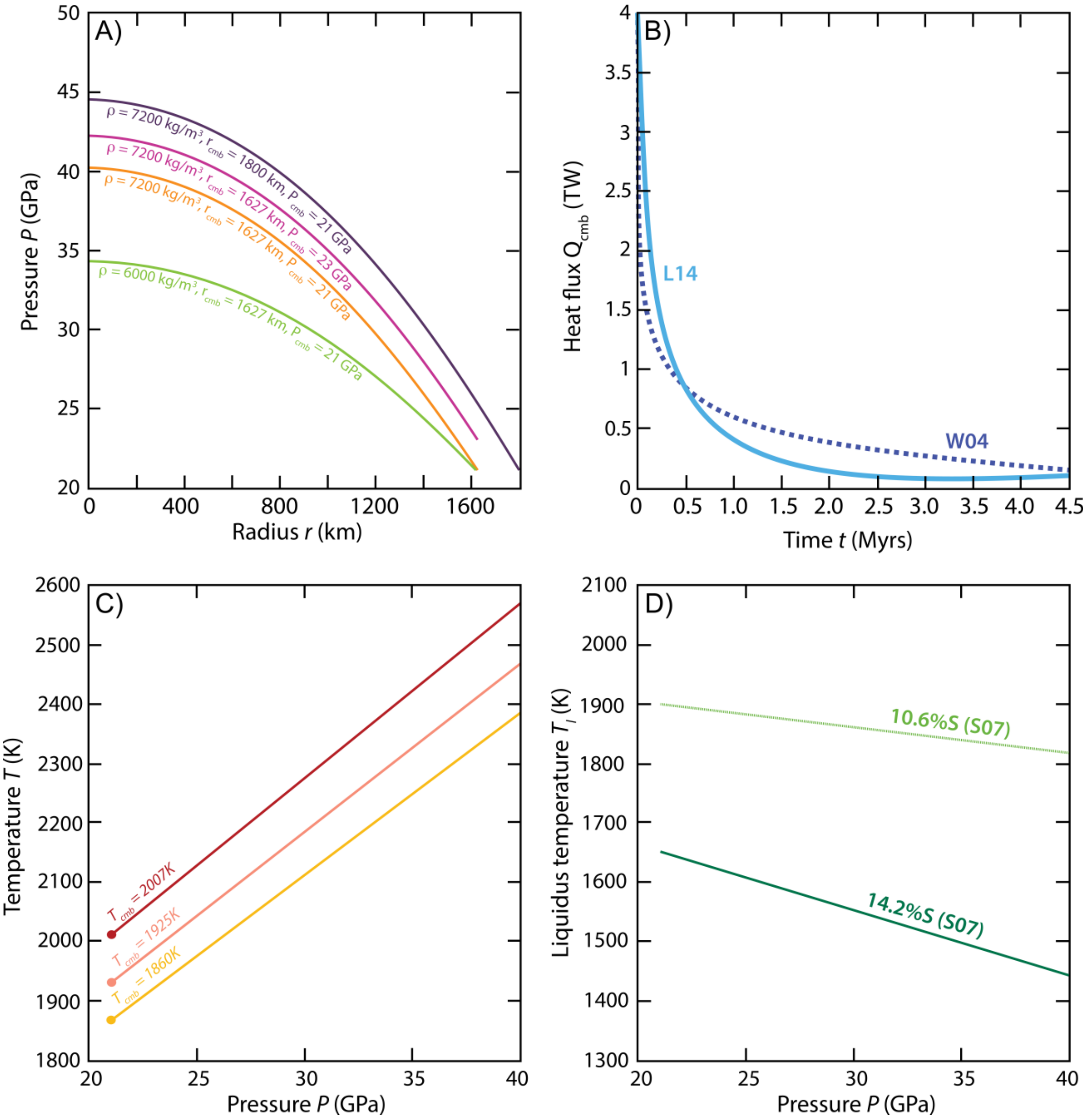

**Figure 2**. Structure of the Martian core. (A) Hydrostatic pressure $P$ plotted against radius $r$ using two end-member values of density $\rho$, CMB radius $r_{cmb}$ and CMB pressure $P_{cmb}$. (B) time-series of CMB heat flow $Q_{cmb}$ obtained as piecewise linear approximations to the original studies. (C) polynomial representation of the adiabatic temperature profile of W04 using 3 different CMB temperatures designed to reflect conditions similar to the present-day. (D) polynomial representations of liquidus temperature. W04 (Williams & Nimmo 2004), L14 (Leone et al. 2014), S07 (Stewart et al. 2007).

**Figure 3**

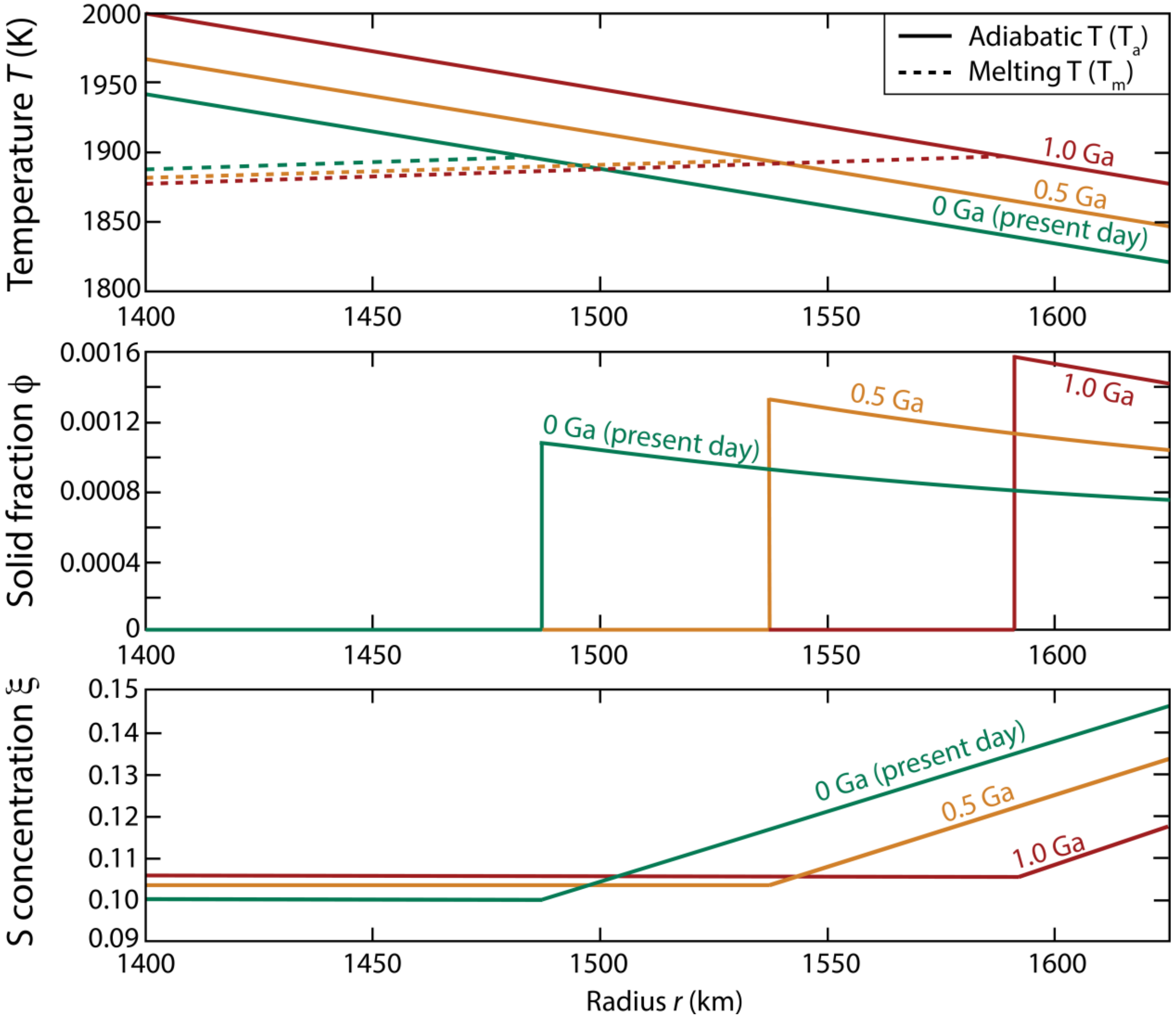

**Figure 3**. Variation of adiabatic temperature $T$, fraction of solid $\phi$ and S concentration $\xi$ as a function of radius $r$ in the upper 227 km of the core. The model uses the default parameters in Table A1. The predicted lower boundary of the snow zone at the present day is at radius $r_s = 1481$ km.

# Figure 4

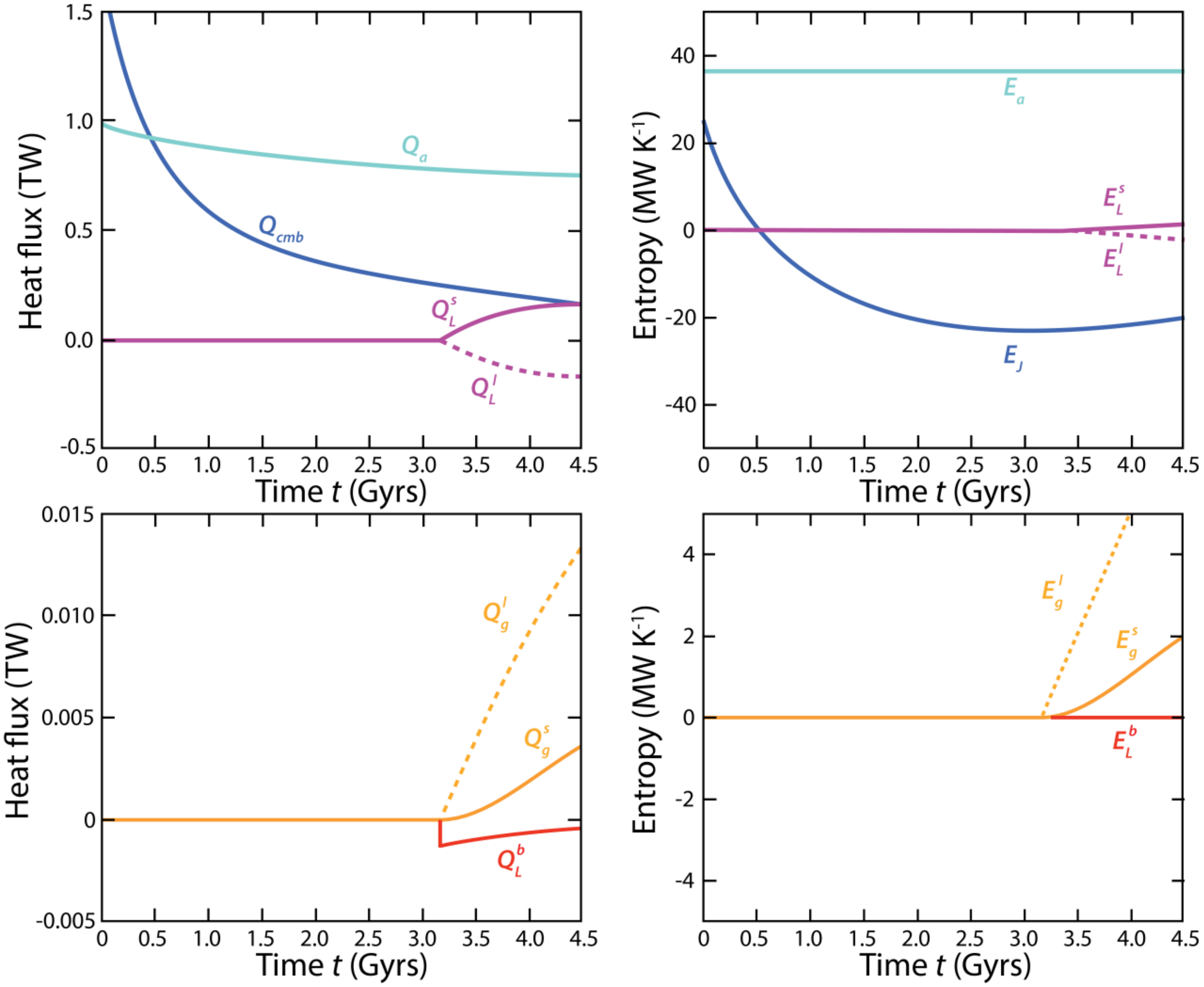

**Figure 4**. Energy (left) and entropy (right) budget for the model shown in Figure 3, which uses the default parameters in Table A1. The range of the abscissa in the bottom row has been reduced compared to the top row so that the contributions of individual terms are visible. Terms are defined in equations (24) and (25).

## Figure 5

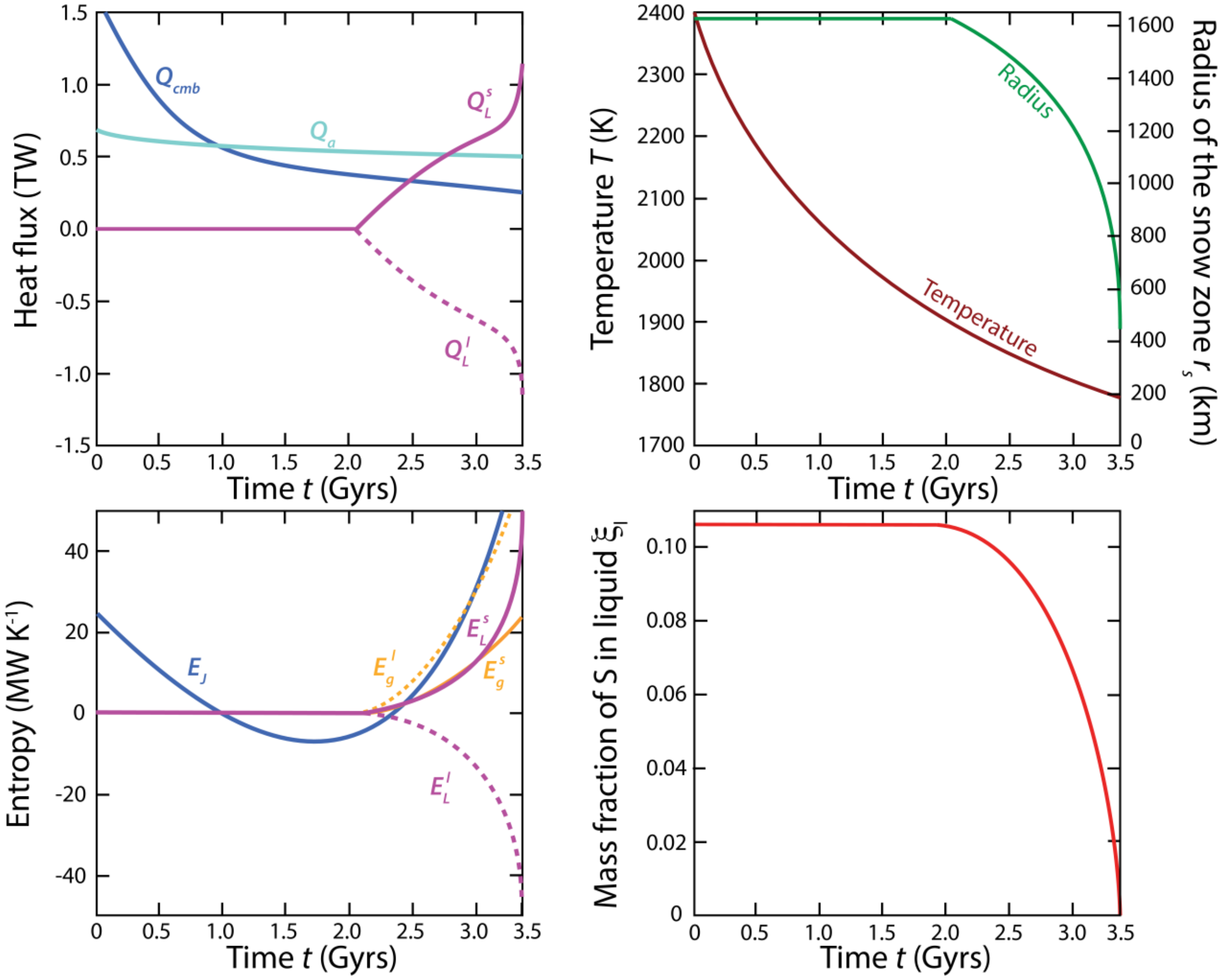

**Figure 5**. Energy budget (top left), entropy budget (bottom left), variation of CMB temperature and snow zone depth (top right) and variation of liquid mass fraction of S (bottom right). The model uses the same parameters as in Figures 3 and 4 except that the core density is set to $\rho = 6000$ kg m$^{-3}$. Small terms in the energy and entropy budgets (see Figure 4) are omitted for clarity.

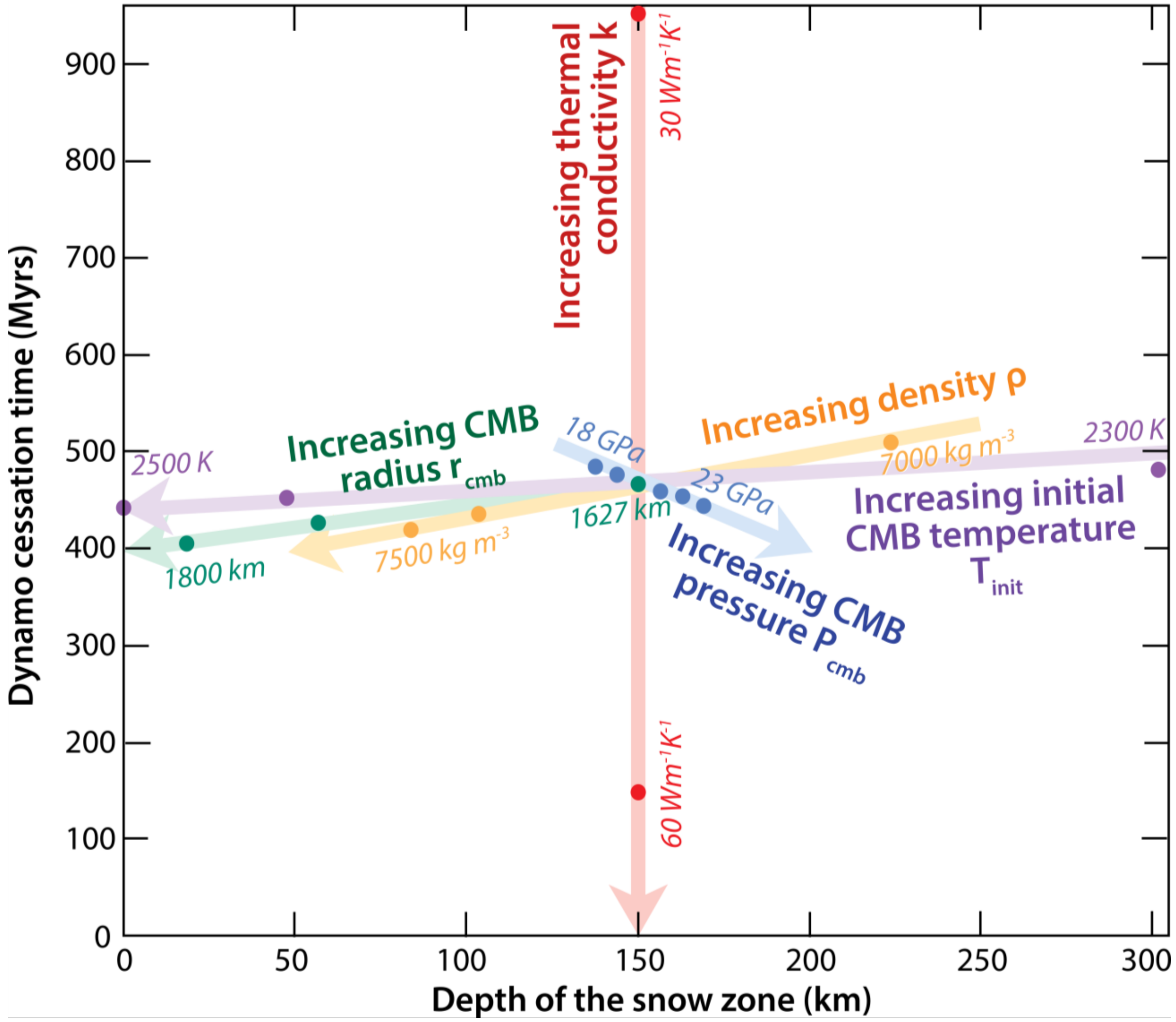

**Figure 6**. Phase diagram illustrating how changes in input parameters alter the predicted snow zone depth $r_s$ (abscissa) and the dynamo cessation time (ordinate). The label for each symbol denotes the single quantity that was changed compared to the default model, which used the parameters highlighted in Table A1 and is shown in Figures 3 and 4. Here $T_{init}$ (K) is the temperature at the CMB at the start of the calculation, $P_{cmb}$ (GPa) is the CMB pressure, $r_{cmb}$ (km) is the CMB radius, $k$ (W m$^{-1}$ K$^{-1}$) is the thermal conductivity and $\rho$ (kg m$^{-3}$) is the density.